\documentclass[twoside]{photon2007}
\usepackage[latin1]{inputenc}
\usepackage[dvips]{graphicx,epsfig,color}
\usepackage{wrapfig,rotating}
\usepackage{amssymb,amsmath,array}

\begin{document}
\title{
%%%%   Paper title goes here  %%%%%%%%%%%%%%
Inclusive photon production \\
and photon-jet correlations \\
in hadronic collisions}
%LATEX Template for Photon2007 Proceedings 
%(adapted from ILC/LCWS07) } 
%% 
%***********************************************************************
% AUTHORS INFORMATION AREA
%***********************************************************************
\author{Antoni Szczurek$^{1,2}$ and Tomasz Pietrycki$^1$
% Optional short acknowledgment: remove next line if non-needed
%\thanks{This is an optional funding source acknowledgment.}
% DO NOT MODIFY THE FOLLOWING '\vspace' ARGUMENT
\vspace{.3cm}\\
% Addresses and institutions (remove "1- " in case of a single institution)
1- Institute of Nuclear Physics PAN, \\
 PL-31-342 Cracow, Poland 
%% Remove the next three lines in case of a single institution
\vspace{.1cm}\\
2- University of Rzesz\'ow, \\
PL-35-959 Rzesz\'ow, Poland 
}
%%***********************************************************************
% END OF AUTHORS INFORMATION AREA
%***********************************************************************

\maketitle

\begin{abstract}
We compare results of $k_t$-factorization approach and next-to-leading order
collinear-factorization approach for photon-jet correlations in $pp$ and
$p \bar p$ collisions at RHIC, Tevatron and LHC energies.
We discuss correlations in azimuthal angle between photon and jet as well as 
correlations in two-dimensional space of photon and jet transverse momenta.
Different unintegrated gluon/parton distributions are used in
the $k_t$-factorization approach. The results depend on UGDF/UPDF used.
The collinear NLO $2 \to 3$ contributions dominate over
$k_t$-factorization cross section at small relative azimuthal
angles as well as for asymmetric transverse momentum configurations.
\end{abstract}

%-----------------------------
\section{Introduction}
%-----------------------------

It was realized relatively early that the transverse momenta of initial
(before a hard process) partons may play an important role in order to
understand the distributions of produced direct photons, especially
at small transverse momenta (see e.g. Ref.~\cite{Owens}). 

The simplest way to include parton transverse momenta is via Gaussian smearing
\cite{Owens,AM04}. This phenomenological approach is not
completely justified theoretically. 

The unintegrated parton distribution functions (UPDFs) are the basic
quantities that take into account explicitly the parton transverse momenta.
The UPDFs have been studied recently in the context of different
high-energy processes (see \cite{LS06} and references therein).
%KMS97,HKSST1,HKSST2,Mariotto,LS04,LS06,LZ05
These works are concentrated
mainly on gluon degrees of fredom which play the dominant role
in many processes at very high energies. At somewhat lower energies
also quark and antiquark degrees of freedom become equally important.
Recently the approach which dynamically includes transverse momenta
of not only gluons but also of quarks and antiquarks was applied
to direct-photon production \cite{KMR_photons,LZ_photon}.

Up to now there is no complete agreement how to include evolution
effects into the building blocks of the high-energy processes --
the unintegrated parton distributions. In Ref.~\cite{PS07_inclusive}
we have discussed in detail a few approaches how
to include transverse momenta of the incoming partons in order to calculate
distributions of direct photons.
In Ref.~\cite{PS07_correlations} we have discussed in addition
photon-jet correlations.

There is recently at RHIC interest in studying hadron-hadron correlations.
The hadron-hadron correlations involve both jet-jet correlations as well
as complicated jet structure.
Recently preliminary data on photon-hadron azimuthal correlation
were presented \cite{RHIC_photon_hadron}.
In principle, such correlations should be easier for theoretical
description as here only one jet enters, at least in leading order pQCD.
On the experimental side, such measurements are more difficult
due to much reduced statistics compared to the dijet studies.

%--------------------------------------------------------------------

\begin{figure}    % Figure 1
\begin{center}
\resizebox{0.40\columnwidth}{!}{%
\includegraphics{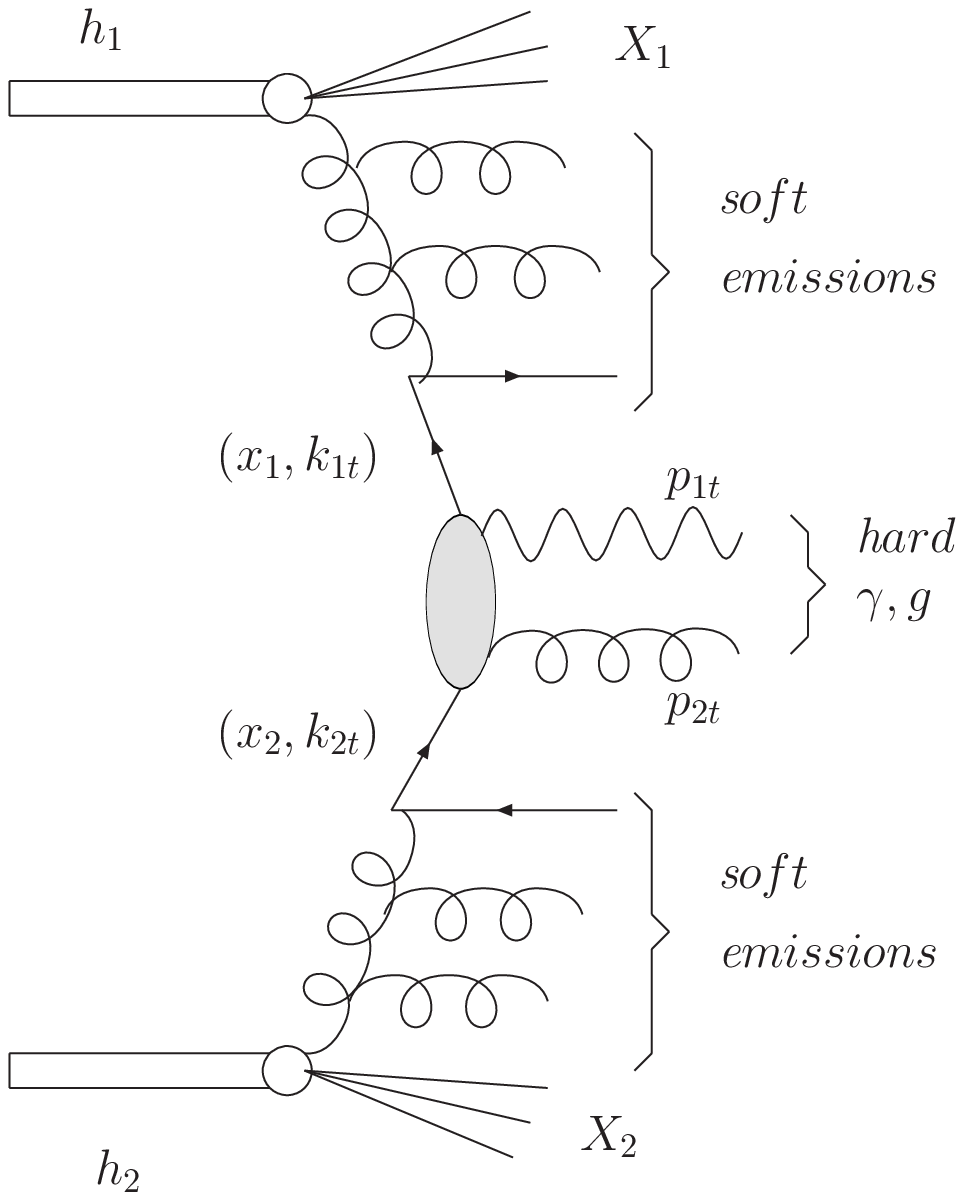} }
\resizebox{0.40\columnwidth}{!}{%
\includegraphics{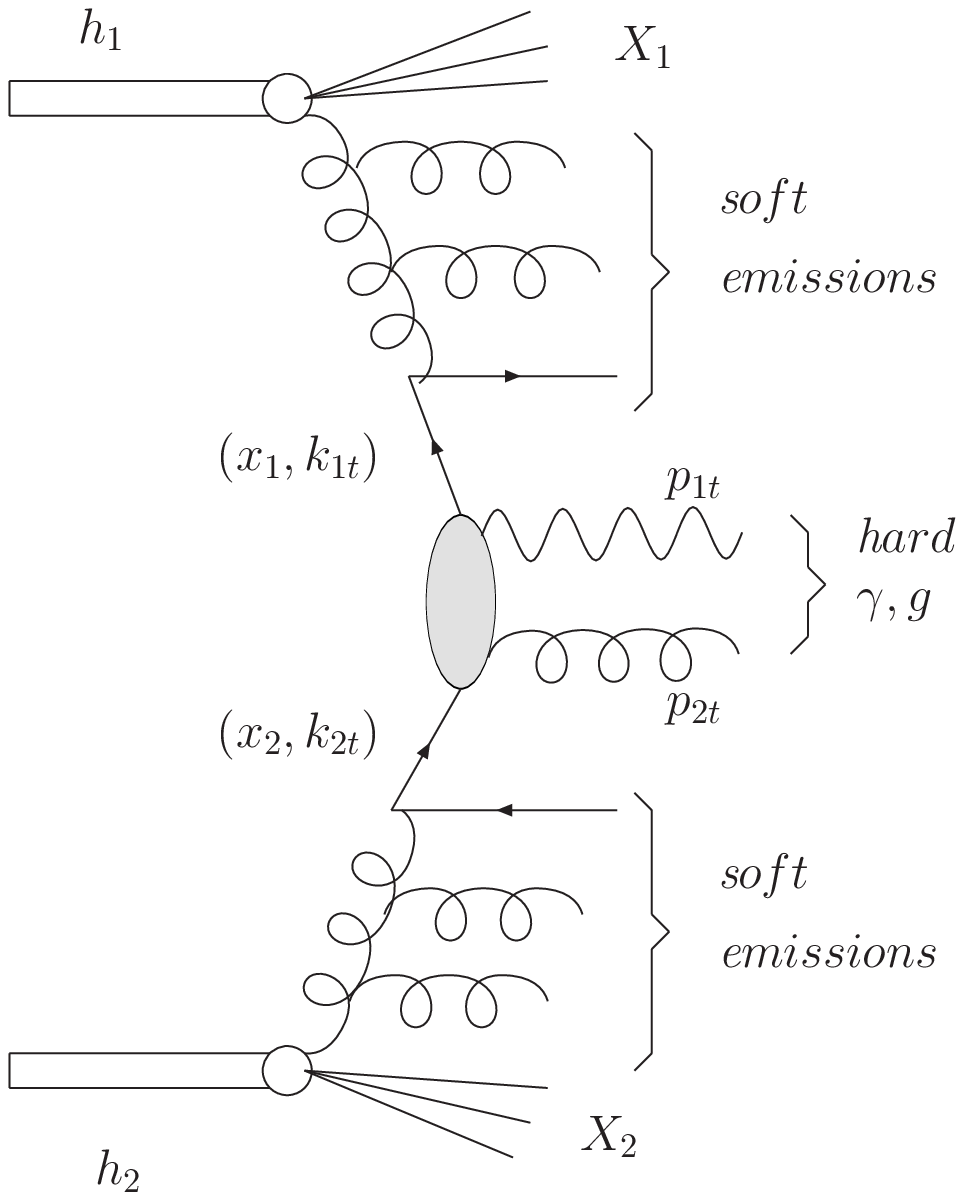} }
\resizebox{0.40\columnwidth}{!}{%
\includegraphics{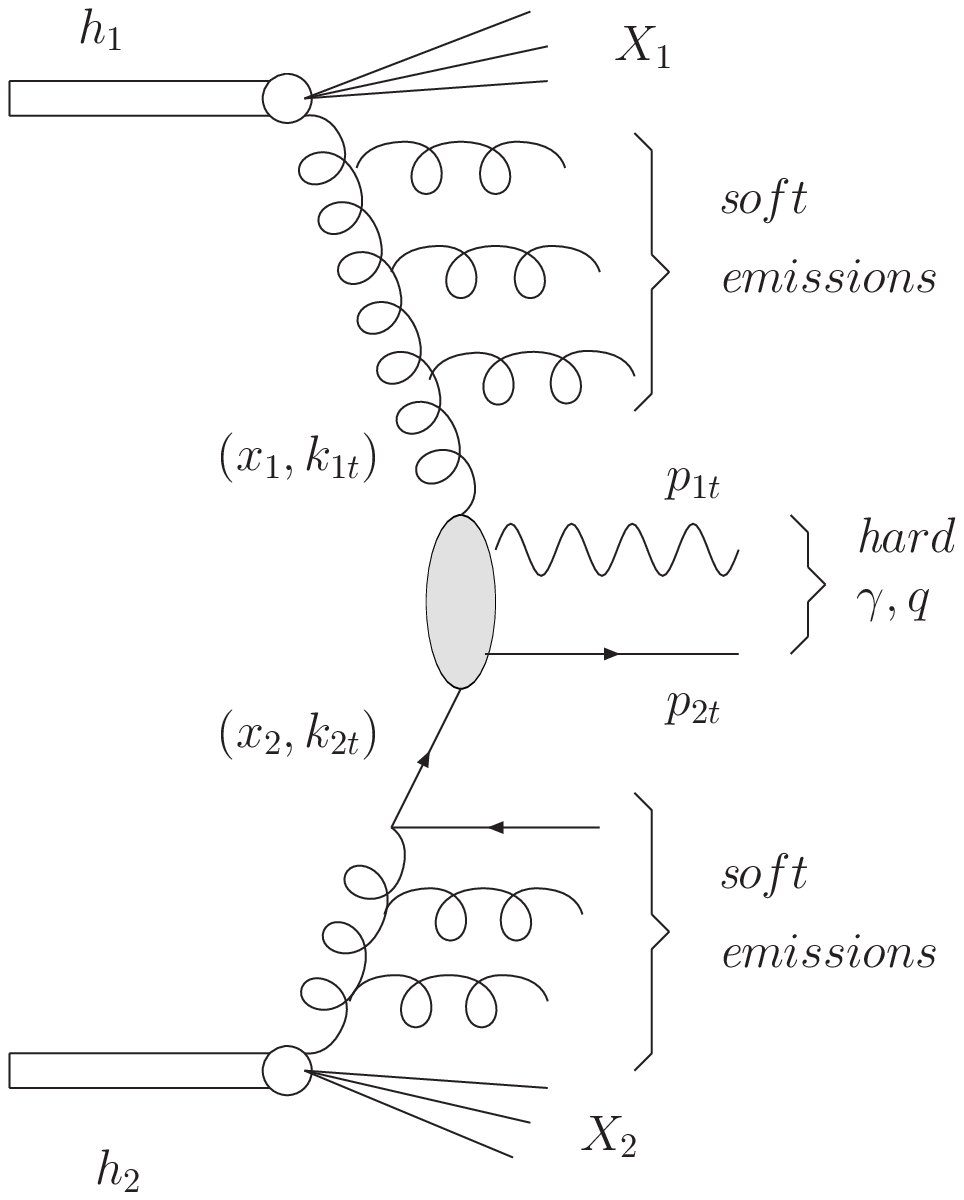} }
\resizebox{0.40\columnwidth}{!}{%
\includegraphics{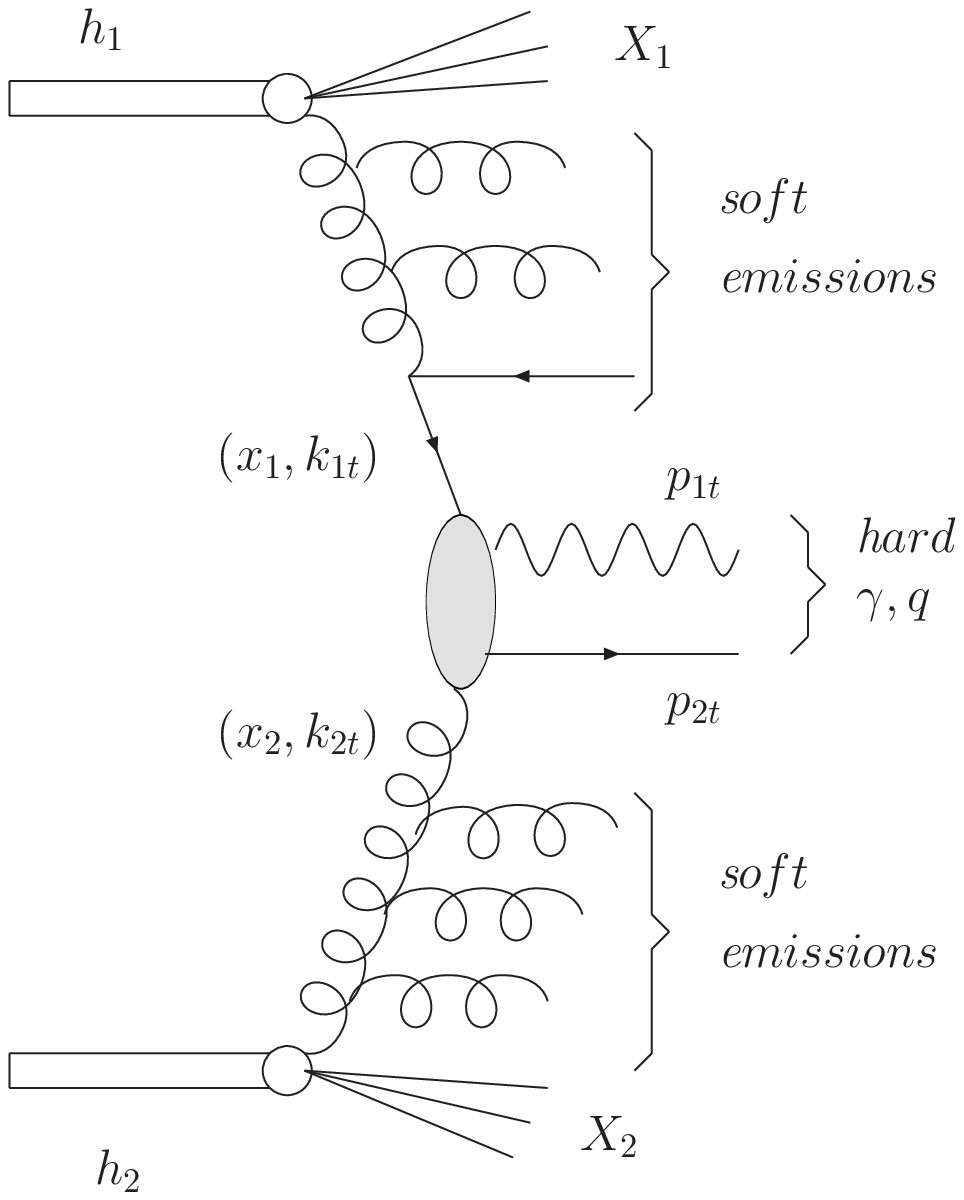} }
\caption{
Basic diagrams for the $k_t$-factorization approach to direct
photon correlations production.}
\label{fig:kt_factorization_photonjet_diagrams}
\end{center}
\end{figure}

%--------------------------------------------------------------------

%---------------------------
\section{Formalism}
%---------------------------

In the $k_t$-factorization approach the cross section for the production
of a pair of photon and parton $(\gamma,l)$ can be written as
\begin{equation}
\begin{split}
&\frac{d\sigma(h_1 h_2 \rightarrow jet(\gamma) jet)}
{d^2p_{1,t}d^2p_{2,t}} 
= \sum_{i,j,l}\\ 
&\int dy_1 dy_2
\frac{d^2 k_{1,t}}{\pi}\frac{d^2 k_{2,t}}{\pi}
\frac{1}{16\pi^2(x_1x_2s)^2}\\
&\overline{|{\cal M}(i j \rightarrow \gamma l)|^2}
\delta^2(\overrightarrow{k}_{1,t}
+\overrightarrow{k}_{2,t}
-\overrightarrow{p}_{1,t}
-\overrightarrow{p}_{2,t})\\
&{\cal F}_i(x_1,k_{1,t}^2){\cal F}_j(x_2,k_{2,t}^2) \; ,
\label{basic_formula}
\end{split}
\end{equation}
where
\begin{equation}
x_1 = \frac{m_{1,t}}{\sqrt{s}}\mathrm{e}^{+y_1} 
    + \frac{m_{2,t}}{\sqrt{s}}\mathrm{e}^{+y_2} \; ,
\end{equation}
\begin{equation}
x_2 = \frac{m_{1,t}}{\sqrt{s}}\mathrm{e}^{-y_1} 
    + \frac{m_{2,t}}{\sqrt{s}}\mathrm{e}^{-y_2} \; ,
\end{equation}
and $m_{1,t}$ and $m_{2,t}$ are so-called transverse masses
defined as $m_{i,t} = \sqrt{p_{i,t}^2+m^2}$, where $m$ is the mass
of a parton.
%In the case of photon-jet correlations there is no sum over $k$ 
%($k = \gamma$).
In the following we shall assume that all partons are massless.
The objects denoted by ${\cal F}_i(x_1,k_{1,t}^2)$ and
${\cal F}_j(x_2,k_{2,t}^2)$ in the equation above are the unintegrated
parton distributions in hadron $h_1$ and $h_2$, respectively.
They are functions of longitudinal momentum fraction and transverse
momentum of the incoming (virtual) parton.

In Fig.\ref{fig:kt_factorization_photonjet_diagrams} we show basic
diagrams included for inclusive photon production and
photon-jet correlations in Refs. \cite{PS07_inclusive,PS07_correlations}.

The formula (\ref{basic_formula}) allows to study different types
of correlations. Here we shall limit to a few examples.
The details concerning unintegrated gluon (parton) distributions
can be found in original publications (see
e.g.\cite{LS06,PS07_inclusive} and references therein).

%======================
\section{Results}
%======================

%-------------------------------------------
\subsection{Inclusive cross sections}
\label{Results:inclusive}
%-------------------------------------------

%In this section we shall present results for RHIC, Tevatron and LHC energies.
In our analysis we use UPDFs from the literature.
There are only two complete sets of UPDFs in the literature
which include not only gluon distributions but also distributions
of quarks and antiquarks: (a) Kwieci\'nski \cite{Kwiecinski}, 
(b) Kimber-Martin-Ryskin \cite{KMR}.
For comparison we shall include also unintegrated parton distributions
obtained from collinear ones by the Gaussian smearing procedure.
Such a procedure is often used in the context of inclusive direct photon
production \cite{Owens,AM04}.
Comparing results obtained with those Gaussian distributions
and the results obtained with the Kwieci\'nski distributions with
nonperturbative Gaussian form factors will allow
to quantify the effect of UPDF evolution as contained in the
Kwieci\'nski evolution equations. What is the hard scale for our process?
In our case the best candidate for the scale 
is the photon and/or jet transverse momentum. Since we are interested
in rather small transverse momenta the evolution length is not
too large and the deviations from initial $k_t$-distributions
(assumed here to be Gaussian) should not be too big.

At high energies one enters into a small-x region, i.e. the region
of a specific dynamics of the QCD emissions. In this region only unintegrated
distributions of gluons exist in the literature. In our case
the dominant contributions come from QCD-Compton
$gluon-quark$ or $quark-gluon$ initiated hard subprocesses. 
This means that we need unintegrated distributions of both
gluons and quarks/antiquarks. In this case we take such UGDFs from the
literature and supplement them by the Gaussian distributions of 
quarks/antiquarks.
%--------------------------------
\begin{figure}[!htb] % Figure 2
\includegraphics[width=5cm]{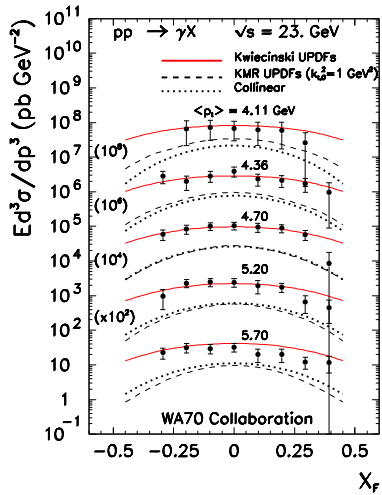}
\caption[*]{
Invariant cross section for direct photons for $\sqrt s$ = 23 GeV 
as a function of Feynman $x_F$ for different bins 
of transverse momenta. In this calculation off-shell 
matrix elements for subprocesses with gluons were used.
The Kwieci\'nski UPDFs were calculated with the factorization 
scale $\mu^2$ = 100 GeV$^2$. The theoretical results 
are compared with the WA70 collaboration data 
\cite{data_WA70} (right panel).
\label{fig:diagrams}
}
\end{figure}
%--------------------------------

In Fig.~\ref{fig:diagrams} we show as an example 
inclusive invariant cross section
as a function of Feynman $x_F$ for several experimental values of
photon transverse momenta as measured by the WA70 collaboration.

It is well known that the collinear approach (dotted line) fails to
describe the low transverse momentum data by a sizeable factor of 4 or
even more. Also the $k_t$-factorization result with the KMR UPDFs
(dashed line) underestimate the low-energy data. In contrast,
the Kwieci\'nski UPDFs (solid line) describe the WA70 collaboration
data \cite{data_WA70} almost perfect.

%-------------------------------
\begin{figure}[!htb] % Figure 3
\begin{center}
\includegraphics[width=5cm]{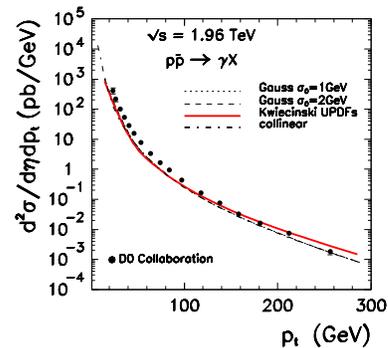}
\caption[*]{
Cross section for direct photons for
$\sqrt s = 1.96$ TeV. In this calculation off-shell matrix element for
gluons were used. The D0 collaboration data were taken from 
Ref.\cite{data_D0_w1960}.
Gaussian smearing ($\sigma_0$ = 1,2 GeV) versus Kwieci\'nski UPDFs. 
\label{fig:inv_cs_w1960}
}
\end{center}
\end{figure}
%\end{widetext}
%-------------------------------

As shown in Ref.\cite{PS07_inclusive} the KMR UPDFs strongly overestimate
the experimental data at large photon transverse momenta. This is especially
visible for proton-antiproton collisions at W = 1.96 TeV when compared
with recent Tevatron (run 2) data \cite{data_D0_w1960}.

In Fig.\ref{fig:inv_cs_w1960} we show an example
of theoretical calculations for the Tevatron energy.
The Kwieci\'nski UPDFs which 
seem to converge to the standard collinear result at large photon
transverse momenta provide relatively good description of the CDF data.

%------------------------------------------
\subsection{Photon-jet correlations}
\label{Results:photon-jet}
%------------------------------------------

\begin{figure}[htb] % Figure 3
\begin{center}
\resizebox{0.48\columnwidth}{!}{%
\includegraphics{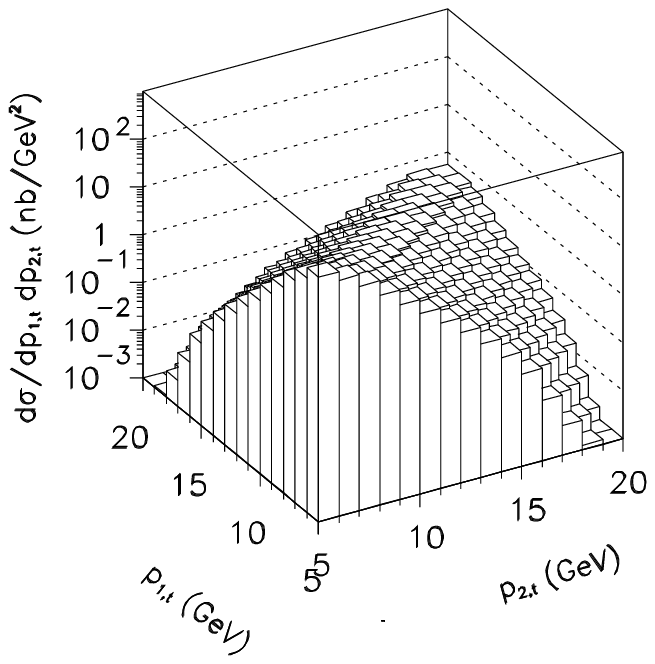}}
\resizebox{0.48\columnwidth}{!}{%
\includegraphics{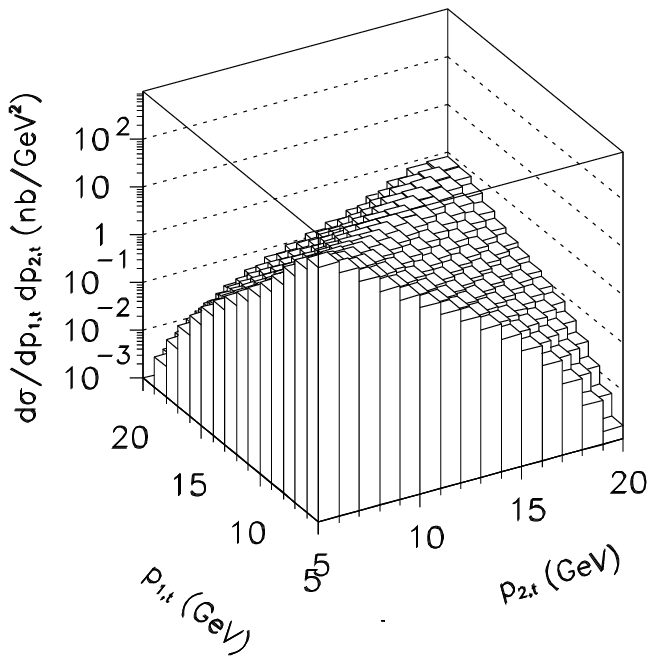}}
\resizebox{0.48\columnwidth}{!}{%
\includegraphics{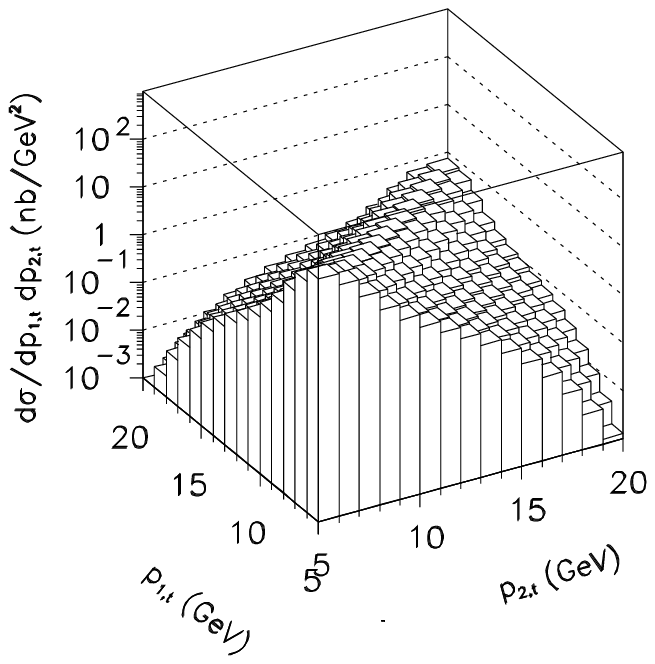}}
\resizebox{0.48\columnwidth}{!}{%
\includegraphics{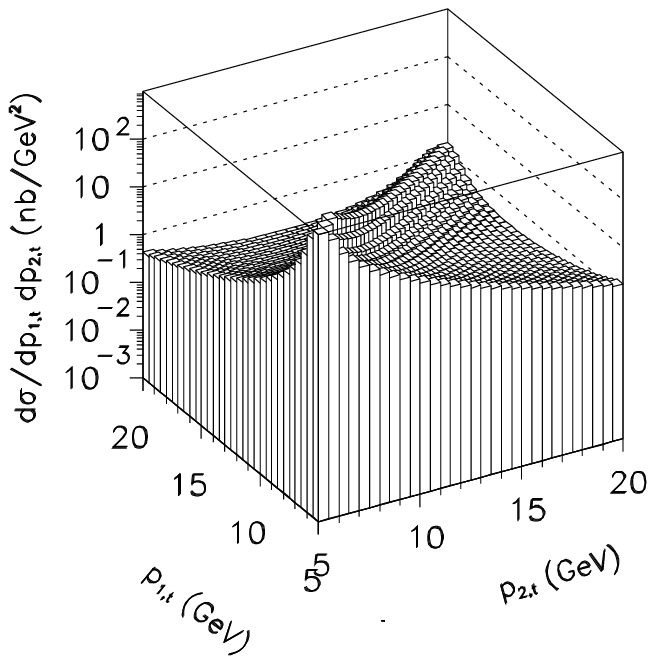}}
\caption{
Transverse momentum distributions $d\sigma/dp_{1,t}dp_{2,t}$
at $\sqrt{s}$ = 1960 GeV and  
for different UPDFs in the $k_t$-factorization approach for
Kwieci\'nski ($b_0$ = 1 GeV$^{-1}$, $\mu^2$ = 100 GeV$^{2}$) (upper left),
BFKL (upper right), KL (lower left) 
and NLO $2 \to 3$ collinear-factorization approach (lower right).
The integration over rapidities from the interval -5 $< y_1, y_2 <$
5 is performed.
}
\label{fig:updfs_ptpt}
\end{center}
\end{figure}

Let us start from presenting our results on the $(p_{1,t},p_{2,t})$ plane.
In Fig.\ref{fig:updfs_ptpt} we show the maps for different 
UPDFs used in the $k_t$-factorization approach as well as for NLO 
collinear-factorization approach for
$p_{1,t}, p_{2,t} \in (5,20)$~GeV and at the Tevatron energy $\sqrt{s} =
1960$~GeV. In the case of the Kwieci\'nski distribution we have taken
$b_0$ = 1 GeV$^{-1}$ for the exponential nonperturbative form factor
and the scale parameter $\mu^2$ = 100 GeV$^2$.
Rather similar distributions are obtained for different UPDFs.
The distribution obtained in the NLO approach differs qualitatively
from those obtained in the $k_t$-factorization approach.
First of all, one can see a sharp ridge along the diagonal $p_{1,t} = p_{2,t}$.
This ridge corresponds to a soft singularity when the unobserved
parton has very small transverse momentum $p_{3,t}$.
At the same time this corresponds to the azimuthal
angle between the photon and the jet being $\phi_{-} = \pi$. Obviously this is
a region which cannot be reliably calculated in collinear pQCD.
There are different practical possibilities to exclude this region from
the calculations \cite{PS07_correlations}.

%------------------------------------------------------------------
%------------------------------------------------------------------

\begin{figure} % Figure 4
\begin{center}
\includegraphics[width=5cm]{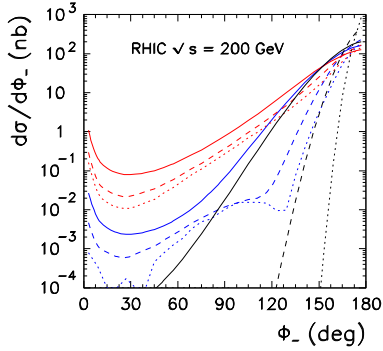}
\caption{
(Color online) Azimuthal angle correlation functions at RHIC
energies for different scales and different values of $b_0$ 
of the Kwieci\'nski distributions.
The solid line is for $b_0$ = 0.5 GeV$^{-1}$, the dashed line is for
$b_0$ = 1 GeV$^{-1}$ and the dotted line is for $b_0$ = 2 GeV$^{-1}$.
Three different values of the scale parameters are shown: 
$\mu^2$ = 0.25, 10, 100 GeV$^2$ (the bigger the scale the bigger
the decorellation effect, different colors on line).
In this calculation  $p_{1,t}, p_{2,t} \in$ (5,20) GeV and
$y_1, y_2 \in$ (-5,5).
}
\label{fig:photon_jet_kwiecinski_scale}
\end{center}
\end{figure}

As discussed in Ref.\cite{PS07_inclusive} the Kwieci\'nski distributions
are very useful to treat both the nonperturbative (intrinsic
nonperturbative transverse momenta)
and the perturbative (QCD broadening due to parton emission) effects on
the same footing.
In Fig.\ref{fig:photon_jet_kwiecinski_scale} we show the effect of
the scale evolution of the Kwieci\'nski UPDFs on the azimuthal angle
correlations between the photon and the associated jet.
We show results for different initial conditions ($b_0$ = 0.5, 1.0, 2.0
GeV$^{-1}$). At the initial scale (fixed here as in the original
GRV \cite{GRV98} to be $\mu^2$ = 0.25 GeV$^2$) there is a sizable
difference of the results for different $b_0$. The difference
becomes less and less pronounced when the scale increases.
At $\mu^2$ = 100 GeV$^2$ the differences practically disappear.
This is due to the fact that the QCD-evolution broadening of
the initial parton transverse momentum distribution is much bigger than
the typical initial nonperturbative transverse momentum scale.

%------------------------------
%-----------

\begin{figure} %  
\begin{center}
\includegraphics[width=5cm]{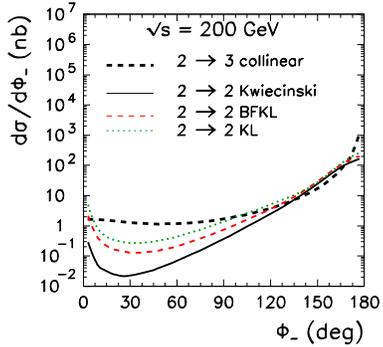}
\caption{
Photon-jet angular azimuthal correlations $d\sigma/d\phi_-$
for proton-(anti)proton collision at $\sqrt{s}$ = 200~GeV  
for different UPDFs in the $k_t$-factorization approach for
the Kwieci\'nski (solid), BFKL (dashed), KL (dotted) UPDFs/UGDFs
and for the NLO collinear-factorization approach (thick dashed).
Here $y_1, y_2 \in (-5,5)$.
}
\label{fig:updfs_phid}
\end{center}
\end{figure}

In Fig.\ref{fig:updfs_phid} we show azimuthal angular
correlations for RHIC.
In this case integration is made over transverse momenta 
$p_{1,t}, p_{2,t} \in (5,20)$~GeV and rapidities $y_1, y_2 \in
(-5,5)$. The standard NLO collinear cross section grows somewhat faster
with energy than the $k_t$-result with unintegrated Kwieci\'nski
parton distributions. This is partially due to approximation made
in calculation of the off-shell matrix elements \cite{PS07_correlations}.

%---------------------------------------------------------------------
%---------------------------------------------------------------------

\begin{figure} % Figure 9
\begin{center}
\includegraphics[width=5cm]{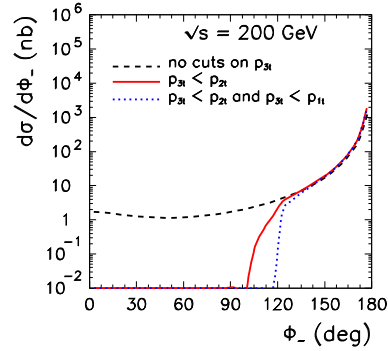}
\caption{
Angular azimuthal correlations for different cuts on the transverse 
momentum of third (unobserved) parton in the NLO collinear-factorization
approach without any extra constraints (dashed), $p_{3,t} < p_{2,t}$ (solid),
$p_{3,t} < p_{2,t}$ and $p_{3,t} < p_{1,t}$ in addition (dotted).
Here $\sqrt{s} = 200$~GeV and $y_1, y_2 \in (-5,5)$.
}
\label{fig:leading_jets_angle}
\end{center}
\end{figure}

Let us consider now some aspects of the standard NLO approach.
%Here 3 jets with transverse momenta $p_{1,t}, p_{2,t}$ and $p_{3,t}$
%are produced \footnote{Jet 1 (with $p_{1,t}$) and jet 2 (with $p_{2,t}$)
%are those which correlations are studied.}.
In Fig.\ref{fig:leading_jets_angle} we show angular azimuthal
correlations for different interrelations between transverse momenta
of outgoing photon and partons:
(a) with no constraints on $p_{3,t}$, (b) the case where $p_{2,t} >
p_{3,t}$ condition (called leading jet condition in the following)
is imposed, (c) $p_{2,t} > p_{3,t}$ and
an additional condition $p_{1,t} > p_{3,t}$.
The results depend significantly on the scenario chosen as can be seen
from the figure. The general pattern is very much the same for different
energies.
The figure demonstrates that only higher-order processes contribute
to the region of small relative azimuthal angles.
We wish to notice that there are no such limitations in the 
$k_t$-factorization approach which implicitly include the higher orders.

%------------------------------
%------------------------------

%------------------------------
\section{Conclusions}
\label{conclusions}
%------------------------------

We have discussed both inclusive production of direct photons
and photon-jet correlations within the $k_t$-factorization approach.
We have concentrated on the region of small transverse momenta
(semi-hard region) where the $k_t$-factorization approach seems to
be the most efficient and theoretically justified tool.
In general, the results of the $k_t$-factorization approach depend
on UGDFs/UPDFs used, i.e. on approximation and assumptions made
in their derivation.

We have obtained very good description of the world data
for the photon single particle distributions with the Kwieci\'nski UPDFs.

An interesting observation has been made for azimuthal angle correlations.
At relatively small transverse momenta ($p_t \sim$ 5--10 GeV)
the $2 \to 2$ subprocesses, not contributing to the correlation
function in the collinear approach, dominate over $2 \to 3$ components.
The latter dominate only at larger transverse momenta, i.e. in
the traditional jet region.
We have calculated correlation observables for different unintegrated parton
distributions from the literature. 

We have discussed the role of the evolution scale of the Kwieci\'nski
UPDFs on the azimuthal correlations. In general, the bigger the scale
the bigger decorrelation in azimuth is observed. When the scale
$\mu^2 \sim p_t^2$(photon) $\sim p_t^2$(associated jet)
(for the  kinematics chosen $\mu^2 \sim$ 100 GeV$^2$) is assumed, 
much bigger decorrelations can be observed than from the standard
Gaussian smearing prescription often used in phenomenological studies.

The correlation function depends strongly on whether it is the
correlation of the
photon and any jet or the correlation of the photon and the leading-jet
which is considered.
In the last case there are regions in azimuth and/or in the two-dimensional
($p_{1,t}, p_{2,t}$) space which cannot be populated in the standard
next-to-leading order approach. In the latter case the $k_t$-factorization
seems to be a useful and efficient tool.

At RHIC one can measure jet-hadron correlations only for not too high
transverse momenta of the trigger photon and of the associated hadron.
This is precisely the semihard region discussed here.
In this case the theoretical calculations would require inclusion of the
fragmentation process. This can be done easily assuming independent parton
fragmentation method.

%---------------------------------------
\section{Acknowledgments}
%---------------------------------------

Antoni Szczurek is indebted to Frederic Kapusta and his colleagues
for nice atmosphere during the conference.

% ****************************************************************************
% BIBLIOGRAPHY AREA
% ****************************************************************************

\begin{footnotesize}
% IF YOU DO NOT USE BIBTEX, USE THE FOLLOWING SAMPLE SCHEME FOR THE REFERENCES
% ----------------------------------------------------------------------------

% ----------------------------------------------------------------------------

\end{footnotesize}

% ****************************************************************************
% END OF BIBLIOGRAPHY AREA
% ****************************************************************************

\end{document}